# Retrieval of Remote Sensing Images Using Colour & Texture Attribute

Priti Maheswary
Research Scholar
Department Of Computer Application
Maulana Azad National Institute of Technology
Bhopal, India
pritimaheshwary@rediffmail.com

Dr. Namita Srivastava
Assistant Professor
Department Of Mathematics
Maulana Azad National Institute of Technology
Bhopal, India
sri.namita@gmail.com

*Abstract* - **Grouping images into semantically meaningful categories using low-level visual feature is a challenging and important problem in content-based image retrieval. The groupings can be used to build effective indices for an image database. Digital image analysis techniques are being used widely in remote sensing assuming that each terrain surface category is characterized with spectral signature observed by remote sensors. Even with the remote sensing images of IRS data, integration of spatial information is expected to assist and to improve the image analysis of remote sensing data. In this paper we present a satellite image retrieval based on a mixture of old fashioned ideas and state of the art learning tools. We have developed a methodology to classify remote sensing images using HSV color features and Haar wavelet texture features and then grouping them on the basis of particular threshold value. The experimental results indicate that the use of color and texture feature extraction is very useful for image retrieval.**

## I. INTRODUCTION

The advent of Digital photography, reduction in cost of mass storage device and use of high-capacity public networks have led to a rapid increase in the use of digital images in various domains such as publishing, media, military and education. The need to store, manage and locate these images has become a challenging task. Generally, there are two main approaches for classifying images: image classification based on keywords and the other one is content based image retrieval. The former technique suffers from the need for manual classification of images, which is simply not practical in a large collection of image. Further incompleteness of a limited set of keyword descriptors may significantly reduce query effectiveness at the time of image retrieval. In latter technique images can be identified by automatic description, which depends on their objective visual content.

Remote Sensing Application images are depicted using spatial distribution of a certain field parameters such as reflectivity of (EM) radiation, emissivity, temperature or some geophysical or topographical elevation. We have designed a system to retrieve similar remote sensing images using some traditional and modern approach.

## II. PREVIOUS WORK

Content Based Image Retrieval is a set of techniques for retrieving semantically relevant images from an image database based on automatically derived image features [1]. The computer must be able to retrieve images from a database without any human assumption on specific domain (such as texture vs. non texture or indoor vs. outdoor).

One of the main tasks for CBIR systems is similarity comparison, extracting feature signatures of every image based on its pixel values and defining rules for comparing images. These features become the image representation for measuring similarity with other images in the database. To compare images the difference of the feature components is calculated.

Early CBIR methods used global feature extraction to obtain the image descriptors. For example, QBIC [2], developed at the IBM Almaden Research Center, extracts several features from each image, namely color, texture and shape features. These descriptors are obtained globally by extracting information on the means of color histograms for color features; global texture information on coarseness, contrast, and direction; and shape features about the curvature, moments invariants, circularity, and eccentricity. Similarly, the Photo-book-system [3], Visual-Seek [4], and Virage [5], use global features to represent image semantics.

The system in [6] attempt to overcome previous method limitations of global based retrieval systems by representing images as collections of regions that may correspond to objects such as flowers, trees, skies, and mountains. This system applies image segmentation [7] to decompose an image into regions, which correspond to physical objects (trees, people, cars, flowers) if the decomposition is ideal. The feature descriptors are extracted on each object instead of global image. Color

(IJCSIS) International Journal of Computer Science and Information Security,
Vol. 4, No. 1 & 2, 2009and texture features are extracted on each pixel that belongs to the object, and each object is described by the average value of these pixel features. In this paper color and texture feature extraction, clustering and similarity matching is used.

### III. METHODOLGY

A system is developed for image retrieval. In this an image database of LISS III sensor is used. LISS III has a spatial resolution of 23m and a swath width of 140 km. Then the query image is taken and images similar to the query images are found on the basis of colour and texture similarity. The three main tasks of the system are:

1. Colour & Texture Feature Extraction.
2. K-means clustering to form groups.
3. Similarity distance computation between the query image and database images.

#### A. Feature Extraction

We have used the approach of Li and Wang [1] and Zhang [9]. The image is partitioned into 4 by 4 blocks, a size that provides a compromise between texture granularity, segmentation coarseness, and computation time. As part of pre-processing, each 4x4 block is replaced by a single block containing the average value of the 4 by 4 block.

To segment an image into objects, six features are extracted from each block. Three features are color features, and the other three are texture features. The HSV color space is selected during color feature extraction due to its ability for easy transformation from RGB to HSV and vice versa. The quantization of HSV can produce a collection of colors that is also compact and complete [6]. These features are {H, S, and V} that are extracted from the RGB colour image.

To obtain the texture features, Haar wavelet transformation is used. The Haar wavelet is discontinuous and resembles a step function. It represents the energy in high frequency bands of the Haar wavelet transform. After a one-level wavelet transform, a 4 by 4 block is decomposed into four frequency bands, each band containing a 2 by 2 matrix of coefficients. Suppose the coefficients in the HL band are $\{c_{k+i}, c_{k,j+1}, c_{k+1,j}, c_{k+1,j+1}\}$. Then, the feature of the block in the HL band is computed as:

$$f = \left( \frac{1}{4} \sum_{i=0}^{1} \sum_{j=0}^{1} c_{k+i,l+j}^2 \right)^{\frac{1}{2}}$$

The other two features are computed similarly in the LH and HH bands. The three features of the block are {HL, LH and LL} [6].

#### B. K-Means Clustering

A cluster is a collection of data objects that are similar to one another with in the same cluster and are dissimilar to the objects in the other clusters. It is the best suited for data mining because of its efficiency in processing large data sets. It is defined as follows:

The k-means algorithm is built upon four basic operations:

1. Selection of the initial k-means for k-clusters.
2. Calculation of the dissimilarity between an object and the mean of a cluster.
3. Allocation of an object of the cluster whose mean is nearest to the object.
4. Re-calculation of the mean of a cluster from the object allocated to it so that the intra cluster dissimilarity is minimized.

After obtaining the six features from all pixels on the image and storing these in an array k-means clustering is performed using Borglet's implementation of K-means clustering [10] to group similar pixel together and form k = 3 clusters. The same procedure is applied on every given image.

The advantage of K-means algorithm is that it works well when clusters are not well separated from each other, which is frequently encountered in images. However, k-means requires the user to specify the initial cluster centers.

#### C. Similarity Matching

Many similarity measures have been developed for image retrieval based on empirical estimates of the feature extraction. Euclidean Distance is used for similarity matching in the present system.

The **Euclidean distance** between two points $P = (p_1, p_2, ......, p_n)$ and $Q = (q_1, q_2, ......, q_n)$, in Euclidean n-space, is defined as:

$$\sqrt{(p_1-q_1)^2 + (p_2-q_2)^2 + \cdots + (p_n-q_n)^2} = \sqrt{\sum_{i=1}^{n}(p_i-q_i)^2}$$

System calculated 6 features of each image objects and then calculates the Euclidean distance of objects of given query image to all three objects of the images in the database.

The distance between two images i.e between query image Q and other image A having three clustered objects as Q/1, Q/2, Q/3 and A/1, A/2, we respectively, have approximated A/3 as follows:

1. Find the Euclidean distance between objects Q/1 to all three objects of A. Let these distances are d1, d2, d3.
2. Find the Euclidean distance between objects Q/2 to all three objects A. Let these distances is d4, d5, and d6.
3. Find the Euclidean distance between object Q/3 to all three objects A. Let these distances are d7, d8, d9.
4. Take M1 as minimum of the three distances d1, d2, and d3.
5. Take M2 as minimum of the three distances d4, d5, and d6.
6. Take M3 as minimum of the three distances d7, d8, and d9.
7. Take the final distance between Q and A as the average of M1, M2, M3 i.e.
   Distance (Q, A) = (M1+M2+M3)/3.

IV. EXPERIMENTAL PLAN

The image retrieval system is implemented using MATLAB image processing tools and statistical tools. For the experiment, system use 12 remote sensing images of urban area obtained from LISS III sensors of 128x128 pixels (Figure 4.1) to perform image retrieval.

*A. Feature Extraction*

Using MATLAB image processing tools, following steps are done for feature extraction:

1. Color and texture features from each pixel are extracted as described in 3.1 (H,S,V for colour and HL, HH, LH for texture).
2. The output of MATLAB code in step one are saved in excel file as an array containing 3 columns of color features and 3 columns of texture features and rows of the total number of pixel on each image.

*B. K-Means Clustering*

Clustering the pixel values obtained from 4.1 using k-means to group similar features together. A sample is shown in table 4.1 of image 1. As can be seen in this table pixel 1 to pixel 17 belongs to cluster 2 and pixel 18 belongs to cluster 1. The results are shown in table 4.1.

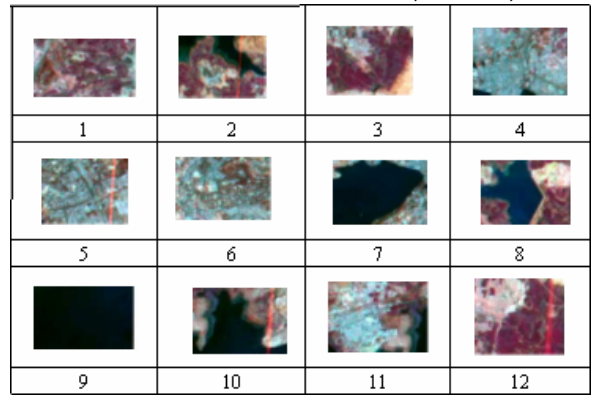

**Fig: 4.1 Images taken as example images**

| PNo | H | S | V | HL | LH | HH | C1 | C2 | C3 |
|---|---|---|---|---|---|---|---|---|---|
| 1 | 0 | 0.12 | 0.44 | -1 | 1 | 0 | 0 | 1 | 0 |
| 2 | 0.7 | 0.05 | 0.66 | 0 | -1 | 1 | 0 | 1 | 0 |
| 3 | 0.86 | 0.08 | 0.76 | 0 | -1 | 1 | 0 | 1 | 0 |
| 4 | 0.61 | 0.11 | 0.8 | -32 | 32 | 32 | 0 | 1 | 0 |
| 5 | 0.7 | 0.09 | 0.87 | -32 | 32 | 32 | 0 | 1 | 0 |
| 6 | 0.66 | 0.08 | 0.9 | -33 | -31.5 | -31.5 | 0 | 1 | 0 |
| 7 | 0.59 | 0.13 | 0.78 | 31 | 72 | 32 | 0 | 1 | 0 |
| 8 | 0.56 | 0.17 | 0.75 | 31 | 72 | 32 | 0 | 1 | 0 |
| 9 | 0.6 | 0.1 | 0.67 | 0 | 0 | 142 | 0 | 1 | 0 |
| 10 | 0.51 | 0.09 | 0.64 | 0 | 0 | 142 | 0 | 1 | 0 |
| 11 | 0.54 | 0.08 | 0.58 | -20 | 21.5 | 162 | 0 | 1 | 0 |
| 12 | 0.43 | 0.2 | 0.49 | -20 | 21.5 | 162 | 0 | 1 | 0 |
| 13 | 0.47 | 0.4 | 0.64 | 20 | 20 | -159 | 1 | 0 | 0 |
| 14 | 0.48 | 0.39 | 0.52 | 20 | 20 | -159 | 1 | 0 | 0 |
| 15 | 0.47 | 0.2 | 0.38 | 1 | -0.5 | 0.5 | 0 | 1 | 0 |
| 16 | 0.75 | 0.04 | 0.42 | 1 | -0.5 | 0.5 | 0 | 1 | 0 |
| 17 | 0.97 | 0.15 | 0.57 | 2 | -0.5 | -0.5 | 0 | 1 | 0 |
| 18 | 0.96 | 0.21 | 0.66 | -4 | -136 | -2.5 | 1 | 0 | 0 |

**Table 4.1 Clustering Result**

|  | 1 | 2 | 3 | 4 | 5 | 6 | 7 | 8 | 9 | 10 | 11 | 12 |
|---|---|---|---|---|---|---|---|---|---|---|---|---|
| 1 | 0 | 0.21 | **0.08** | 0.18 | 0.17 | 0.19 | 0.17 | 0.17 | **0.27** | 0.16 | 0.17 | 0.21 |

**Table 4.2: Distance between image 1 and all other images**

*C. Similarity Matching*

Images similar to the query image are retrieved. The distance of image 1 is calculated from all the images using Euclidean distance. The final distance between the query image1 and the other entire images in database is shown in table 4.2.

Distance of image 1 to image 2 is 0.211575 while the distance of image 1 to image 7 is 0.174309.

Consider Image 4 as Query, the table 4.3 shows the distances (threshold between 0 and 0.1) with the closest images. As it is clear from the table that image 11 is closest to image 4.

| Image 4 | Image 5 | Image 6 | Image 10 | Image 11 |
|---|---|---|---|---|
| 0 | 0.079 | 0.097 | **0.11** | **0.077** |

**Table 4.3: Distance between image 4 and all other similar images.**

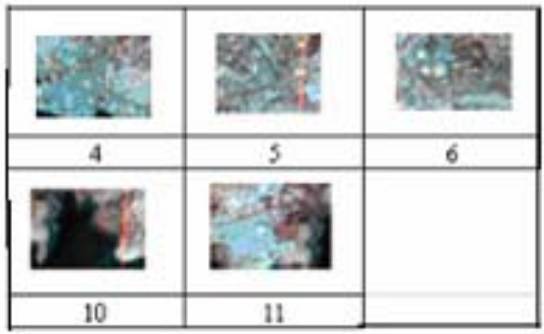

**Fig: 4.2: Similar images of query image 4.**

Consider Image 3 as Query, table 4.4 shows the distances (between 0 and 0.2) with the closest images. It is clear from the table that image 3 is closest to image 1.

| Image 3 | Image 1 | Image 2 | Image 12 |
|---|---|---|---|
| 0 | 0.08195 | 0.19374 | 0.173131 |

**Table 4.4: Distance between image 3 and all other similar images**

## 5. CONCLUSION

For retrieving similar images to a given query image we have tried to perform the segmentation of images using color & texture feature and then clustering of image features and finally calculate the similarity distance. Color Feature Extraction is done by HSV color space and texture feature extraction is done by haar wavelet transformation. Grouping of objects in the data is performed using K-means clustering algorithm. Similarity matching of images is based on Euclidean Distance. We get fruitful results on the example images used in the experiments. We can use this technique for mining similar images based on content and knowledge base for finding vegetation or water or building areas.

## 6. REFERENCES

[1] Li, J., Wang, J. Z. and Wiederhold, G., "Integrated Region Matching for Image Retrieval," *ACM Multimedia*, 2000, p. 147-156.

[2] Flickner, M., Sawhney, H., Niblack, W., Ashley, J., Huang, Q., Dom, B., Gorkani, M., Hafner, J., Lee, D., Petkovic, D., Steele, D. and Yanker, P., "Query by image and video content: The QBIC system," *IEEE Computer*, **28(9)**, 1995,pp.23-32

[3] Pentland, A., Picard, R. and Sclaroff S., "Photobook: Contentbased manipulation of image databases", *International Journal of Computer Vision*, **18(3)**, 1996, pp.233–254

[4] Smith, J.R., and Chang, S.F., "Single color extraction and image query," *In Proceeding IEEE International Conference on Image Processing*, 1997, pp. 528–531

[5] Gupta, A., and Jain, R., "Visual information retrieval," *Comm. Assoc. Comp. Mach.*, **40(5)**, 1997, pp. 70–79

[6] Eka Aulia, "Heirarchical Indexing for Region based image retrieval", A thesis Submitted to the Graduate Faculty of the Louisiana State University and Agricultural and Mechanical College.

[7] Shi, J., and Malik, J., "Normalized Cuts and Image Segmentation," Proceedings *Computer Vision and Pattern Recognition*, June, 1997, pp. 731-737

[8] Smith, J., "Color for Image Retrieval", *Image Databases: Search and Retrieval of Digital Imagery, John Wiley & Sons*, New York, 2001, pp.285-311

[9] Zhang, R. and Zhang, Z., (2002), "A Clustering Based Approach to Efficient Image Retrieval," *Proceedings of the 14th IEEE International Conference on Tools with Artificial Intelligence*, pp. 339

[10]　http://fuzzy.cs.uni-magdeburg.de/~borgelt/software for kmeans clustering software.